\documentclass[twocolumn,superscriptaddress,showpacs,prl]{revtex4-1}    
\usepackage{graphicx,amsmath,mathrsfs}
\usepackage{amssymb}
\usepackage{amsthm}
\usepackage{bm}
\usepackage{subfigure,xcolor}   

\def\bc{\begin{center}}
\def\ec{\end{center}}

\newcommand{\bs}[1]{\boldsymbol{#1}}

\newcommand{\up}{\uparrow}
\newcommand{\dw}{\downarrow}
\newcommand{\x}{\text{x}}
\newcommand{\y}{\text{y}}
\newcommand{\z}{\text{z}}
\newcommand{\F}{\text{F}}
\newcommand{\zz}{\text{Z}}
\newcommand{\SO}{\text{so}}
\newcommand{\R}{\text{R}}
\renewcommand{\L}{\text{L}}
\renewcommand{\i}{\text{i}}
\renewcommand{\c}{\text{c}}
\newcommand{\s}{\text{s}}

\def\ie{\emph{i.e.},\ }

\begin{document}
\title{Coupled wire construction of chiral spin liquids}
\author{Tobias Meng} \affiliation{Institut f\"ur Theoretische Physik,
  Technische Universit\"at Dresden, 01062 Dresden, Germany}
\author{Titus Neupert} \affiliation{Princeton Center for Theoretical
  Science, Princeton University, Princeton, New Jersey 08544, USA}
\author{Martin Greiter} \affiliation{Institute for Theoretical
  Physics, University of W\"urzburg, 97074 W\"urzburg, Germany}
\author{Ronny Thomale} \affiliation{Institute for Theoretical Physics,
  University of W\"urzburg, 97074 W\"urzburg, Germany}


\begin{abstract}
  We develop a coupled wire construction of chiral spin liquids. The
  starting point are individual wires of electrons in the Mott regime
  that are subject to a Zeeman field and Rashba spin-orbit
  coupling. Suitable spin-flip couplings between the wires yield an
  Abelian chiral spin liquid state which supports spinon excitations
  above a bulk gap, and chiral edge states. The approach generalizes to
  non-Abelian chiral spin liquids at level $k$ with parafermionic edge
  states.
\end{abstract}



\maketitle

\emph{Introduction.}---The experimental discovery~\cite{tsui-82prl1559}
and conceptual understanding~\cite{laughlin83prl1395} of the fractional
quantum Hall effect (FQHE) had a tremendous impact on contemporary
research of strongly correlated electron systems. In particular, it
triggered interest in topologically ordered quantum states of matter,
which since then have persisted as a predominant focus. Following up on
an idea by D.~H.~Lee, Kalmeyer and
Laughlin~\cite{kalmeyer-87prl2095,kalmeyer-89prb11879} proposed the
chiral spin liquid~\cite{wen-89prb11413,laughlin-90prb664} (CSL) as a fractionally quantized Hall liquid for
bosonic spin flip operators acting on a spin-polarized reference
state. Fractionalization of charge for FQHE relates
to fractionalization of spin for the CSL, which supports $S=1/2$ spinons obeying half-Fermi statistics~\cite{wilczek90}. The CSL has been an invaluable seed for new concepts such as
topological order~\cite{wen89prb7387}, providing a direct perspective
on the fundamental relations between FQHE, spin liquids, and
superconductivity~\cite{wen-89prb11413,Read-00prb10267}. Despite its high relevance as a paradigm formulated via wave
functions, the first Hamiltonian for which the CSL is the (aside from
topological degeneracies) unique ground state was only identified two
decades after the liquid had been
proposed~\cite{schroeter-07prl097202}.  The approach was subsequently
expanded to yield different classes of such trial
Hamiltonians~\cite{thomale-09prb104406}, where the latest and more
generic versions are more short-ranged than the initial microscopic
models: they involve 2-body and 3-body spin interactions which can be
deduced from the explicit construction of appropriate annihilation
operators~\cite{greiter-14prb165125} or null operators in conformal
field theory~\cite{nielsen-12prl257206}.  In particular, non-Abelian
chiral spin liquids with level $k$ parafermionic spin excitations have
been proposed~\cite{greiter-09prl207203,greiter-14prb165125}, which
nurture the hope for alternative scenarios of topological quantum
computation in frustrated magnets and Mott regimes of alkaline earth
atoms deposited in optical
lattices~\cite{hermele-09prl135301,nielsen-13nc3864}.

Since their discovery, CSLs have been appreciated as a realisation of
a bosonic Laughlin state at Landau level filling fraction $\nu=1/2$ on
a spin lattice. Naturally, the CSL of
Refs.~\onlinecite{kalmeyer-87prl2095,kalmeyer-89prb11879} can be
defined on any lattice~\cite{zou-89prb11424,yang-93prl2641}, which
becomes mathematically transparent via the generalized Perelomov
identity~\cite{greiter-12prb155145} for lattices with a primitive unit
cell. Some of these motifs have later reappeared in the field of
fractional Chern
insulators~\cite{tang-11prl236802,sun-11prl236803,neupert-11prl236804}. As
of today, several promising CSL scenarios with broken SU(2) spin
symmetry have been proposed, while analytic wave functions are not
known in these cases. The most important one is the Kitaev model on
the decorated honeycomb lattice~\cite{yao-07prl247203}, which can be
solved exactly by a mapping to Majorana fermions. In addition, recent
large-scale numerical studies are interpreted in favor of a CSL regime
in models of broken~\cite{gong-15prb075112} and
conserved~\cite{messio-12prl207204,bauer-14nc5137,hu-15prb041124,laeuchli-arxiv}
SU(2) spin symmetry on the kagome lattice, where competing magnetic
order is sufficiently frustrated. From the viewpoint of symmetry
classification, SU(2) symmetry is not a characteristic feature of
CSLs. In contrast, parity (P) and time-reversal (T) symmetry are
necessarily broken in CSLs, and they support (gaped) spinon
excitations in an otherwise featureless fluid.

In this Letter, we develop a coupled wire construction (CWC) of CSL
states. The CWC for topologically ordered quantum states of matter
originates from the pioneering work by Kane and collaborators on
deriving a scenario of FQH states from suitably chosen many-particle
couplings in a set of coupled quantum
wires~\cite{kane-02prl036401}. Important preceding work has been on
sliding Luttinger liquid phases, which already installed the notion of
using the magnetic field as a way to favorably tune the scaling
dimension of inter-wire couplings~\cite{sondhi-01prb054430}.
Recently, the CWC of two-dimensional systems has been employed in
various
contexts~\cite{lu-12prb125119,vaezi-14prl236804,seroussi-14prb104523,klinovaja-14epjb87,meng-14epjb203,sagi-14prb201102,meng-14prb235425,klino_add_1,klino_add_2}
including a derivation of the periodic table of integer and fractional
fermionic topological phases~\cite{neupert-14prb205101}.  In a way,
the CWC of Abelian and non-Abelian CSLs reported in this paper
completes the program previously pursued for the CWC of Read-Rezayi
states in the FQHE~\cite{Read-99prb8084,teo-14prb085101} and their
superconducting analogues~\cite{mong-14prx011036,vaezi14prx031009}. As
such, the CWC provides a fruitful perspective on a broad range of
non-Abelian topological quantum states of matter.



\begin{figure} \centering
\includegraphics[width=0.75\columnwidth]{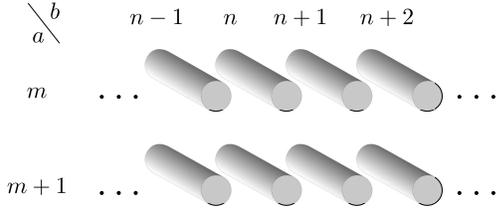}\\
  \caption{Wires $b=n-1,\ldots,n+2$ of layers $a=m$ and $m+1$ of the
multi layer array of coupled wires.}
  \label{fig:array}
\end{figure}

\emph{Model of coupled wires.}---We analyze a $k$-fold stacked array
of $N$ quantum wires, as shown in Fig.~\ref{fig:array}. We label the
wires by $(a,b)$, where $a=1,\ldots,k$ is the layer (or flavour)
index, and $b=1,\ldots,N$ is the wire index within each layer. Each
wire is modeled by a cosine band of spinful electrons, subject to a
Zeeman field and spin-orbit coupling.  The couplings are constant within a given wire, but depend on the wire index $b$.
The four Fermi points of right (R) and left
(L) moving electrons of spin $\sigma=\,\up,\dw$ in wire $(ab)$ reside
at momenta
\begin{align}
  k_{\text{F}r\sigma}^{(b)}&=k_\text{F}
  +\sigma k_{\zz}^{(b)}+r\sigma k_{\SO}^{(b)},\label{eq:fermi_points}
\end{align} 
where $r=\R,\L$. We identify $\R,\up\,\equiv+$ and $\L,\dw\,\equiv-$.
Here, $k_{\F}$ denotes the bare Fermi momentum, and $k_{\zz}^{(b)}$
and $k_{\SO}^{(b)}$ the momentum shifts due to the Zeeman field and
the spin-orbit coupling, respectively.

Throughout this paper, we treat the system in (Abelian)
bosonization~\cite{giamarchi04}. Linearising the spectrum around the
Fermi points, the electron annihilation operators
$c_{\sigma}^{(ab)}(x)$ can be decomposed into right and left
moving modes, \begin{align}
  c_{\sigma}^{(ab)}(x) 
  =e^{\i k_{\text{F}\R\sigma}^{(b)} x}R_{\sigma}^{(ab)}(x)
  +e^{-\i k_{\text{F}\L\sigma}^{(b)} x}L_{\sigma}^{(ab)}(x),
\end{align}
which we bosonize as 
\begin{align}
  r_{\sigma}^{(ab)}(x)=
  \frac{U_{r\sigma}^{(ab)}}{\sqrt{2\pi\alpha}}\,
  e^{-\i [r\phi_\sigma^{(ab)}(x)-\theta_\sigma^{(ab)}(x)]},
\end{align}
where $U_{r\sigma}^{(ab)}$ is a Klein factor, and $\alpha$ the
short distance cutoff of the theory. The bosonic fields obey
\begin{align}
  \bigl[\phi_{\sigma}^{(ab)}(x),\theta_{\sigma'}^{(ab)}(x')\bigr]
  =-\delta_{aa'}\delta_{bb'}\delta_{\sigma\sigma'}
  \frac{\i\pi}{2}\text{sgn}(x-x').
\end{align}
With these definitions, the long wavelength density fluctuations of
spin $\sigma$ electrons in wire $(ab)$ are given by
$\rho_{\sigma}^{(ab)}(x)
=-\frac{1}{\pi}\partial_x\phi_{\sigma}^{(ab)}(x)$. In terms of the
charge $\c$ and spin $\s$ modes of the wires, defined as
\begin{align}
  \phi_{\c /\s}^{(ab)} =\frac{1}{\sqrt{2}}
  (\phi_{\uparrow}^{(ab)}\pm\phi_{\downarrow}^{(ab)}),\ \ 
  \theta_{\c /\s}^{(ab)} =\frac{1}{\sqrt{2}}
  (\theta_{\uparrow}^{(ab)}\pm\theta_{\downarrow}^{(ab)}),
\end{align}
the Hamiltonian density of an individual wire reads
\begin{align} 
  \mathcal{H}_0^{(ab)}(x)
  &=\Psi_{ab}^T(x) V_{ab} \Psi_{ab}(x)\nonumber\\[2pt]
  &+\frac{g_{3}}{(2\pi\alpha)^2}\,
  e^{-\i\sum_{r\sigma}k_{\text{F}r\sigma}^{(b)}x}\,
  e^{\i\sqrt{8}\phi_c^{(ab)}(x)}+\text{H.c.}\nonumber\\
  &+\frac{g_{1\perp}}{(2\pi\alpha)^2}\,
  e^{-\i\sum_{r\sigma}\sigma k_{\text{F}r\sigma}^{(b)}x}\,
  e^{\i\sqrt{8}\phi_{\s}^{(ab)}(x)}+\text{H.c.},\label{eq:ham_single_wire}
\end{align} 
where the $(4\times4)$-matrix $V_{ab}$ depends on the Fermi velocity
and forward scattering interactions, and
\mbox{$\Psi_{ab}=(\partial_x\phi_c^{(ab)},\partial_x\phi_{\s}^{(ab)},
\partial_x\theta_c^{(ab)},\partial_x\theta_{\s}^{(ab)})^T$}. In addition to the quadratic part, the Hamiltonian also contains a Mott term $\sim g_3$,
and a backscattering term $\sim g_{1\perp}$. In a clean system, these terms contribute only when the oscillating prefactors vanish.




\emph{Mott gap and spin flip operators.}---For the construction of a
spin liquid, we gap out the charge sector in each wire by tuning it
into the Mott regime. Equation~\ref{eq:ham_single_wire} implies that
this regime can be reached for
$\sum_{r\sigma}k_{\text{F}r\sigma}^{(b)} = 2\pi/\alpha_0$, where
$\alpha_0$ is the lattice constant (taken identical in all wires). We
thus demand $k_\text{F}= \pi /2\alpha_0$.  At the same time, the spin
sectors should remain gapless if there are no inter-wire
couplings. This either requires the sine-Gordon term {$\sim g_{1\perp}$} to
be irrelevant in the sense of the renormalization group (RG) (or less
relevant than the couplings stabilizing the spin liquids, which are
discussed below), or $\sum_{r\sigma}\sigma k_{\text{F}r\sigma}^{(ab)}
= 4k_{\zz}^{(ab)} \neq 0$. We thus apply a Zeeman field in all wires.

In the Mott phase, the single wire Hamiltonian densities
$\mathcal{H}_0^{(ab)}$ pin the fields $\phi_c^{(ab)}$ to values
$\phi_c^{(ab)} \approx \langle\phi_c^{(ab)}\rangle$.  
(The value of $\langle\phi_c^{(ab)}\rangle$ depends on the sign
convention for the ordering of the Klein factors.  Taking them to be
Majorana fermions with $U_{r\sigma}^\dagger=U_{r\sigma}$ and
$U_{r\sigma}^2=1$, we choose $U_{\L\up}U_{\L\dw}U_{\R\up}U_{\R\dw}=1$
on each wire $(ab)$.  A Hubbard interaction $U c_\up^\dagger
c_\up^{\phantom{\dagger}} c_\dw^\dagger c_\dw^{\phantom{\dagger}}$
then generates $g_3=-U<0$, which implies we may take
$\langle\phi_c^{(ab)}\rangle=0$.)

In the Mott phase, the remaining local degrees of freedom are spin flip operators
\begin{equation}
\begin{split}
  S_{ab}^+ &= c_{\uparrow}^{(ab)}{}^\dagger
  c_{\downarrow}^{(ab)}\\
  &=\sum_{r=\R,\L} 
  \frac{U_{r\uparrow}^{(ab)}{}^\dagger
    U_{r\downarrow}^{(ab)}}{2\pi\alpha}\, 
  e^{-r\i k_{1r}^{(b)}x}\, e^{\i\sqrt{2}(r\phi_{\s}^{(ab)}-\theta_{\s}^{(ab)})}\\
  &+\sum_{r=\R,\L}
  \frac{U_{r\uparrow}^{(ab)}{}^\dagger
  U_{-r\downarrow}^{(ab)}}{2\pi\alpha} 
  e^{-r\i k_{2r}^{(b)}x}\,e^{\i\sqrt{2}(r\phi_{\c}^{(ab)}-\theta_{\s}^{(ab)})},\label{eq:sab}
\end{split}
\end{equation} 
where
\begin{align} 
  k_{1r}^{(b)} &\equiv  2(k_\zz^{(b)}+r k_{\SO}^{(b)}),\ \ 
  k_{2r}^{(b)} \equiv  2(k_\text{F}+r k_{\SO}^{(b)}).\label{eq:k}
\end{align}




\emph{Abelian chiral spin liquid.}---The Abelian CSL only requires a
single layer, or flavour, of wires ($k=1$). Similar to the wire
construction of quantum Hall
states~\cite{kane-02prl036401,teo-14prb085101}, we couple right movers
in wire $b$ to left movers in wire $b+1$, but not in wire $b-1$. Such
a coupling breaks time reversal symmetry T as well as two-dimensional
parity P (which we take as $x\to x$, $y\to -y$ along and transverse to
the wires, respectively), but is PT invariant. (The conservation of
PT, however, is no prerequisite for a CSL in the sense of a
symmetry-protected topological (SPT) phase~\cite{lu-12prb125119}.)

In the construction of Kalmeyer and
Laughlin~\cite{kalmeyer-87prl2095,kalmeyer-89prb11879} (KL), P and T
are violated through the fictitious magnetic field used to stabilise
the $m=2$ Laughlin state, which is then projected to describe spin
flips on a lattice commensurable with the magnetic field (one Dirac
flux quantum per unit cell of the square lattice). While readily
implemented in a language of wave functions, this method of obtaining
a CSL state in two steps---writing it out in the continuum and then
projecting it onto the lattice---is not available in a description in
terms of Hamiltonians, such as our CWC. In all known
constructions~\cite{schroeter-07prl097202,thomale-09prb104406,greiter-14prb165125,nielsen-12prl257206}
of parent Hamiltonians of the KL state, P and T violation is
implemented through a three-spin interaction of the form
$\bs{S}_i(\bs{S}_j\times\bs{S}_k)$, where $i$, $j$, and $k$ are three
lattice sites on a plaquette.

A term of this form, however, is not easily implemented as a coupling
between wires.  A simpler and more elegant way to obtain a CSL,
{\it i.e.}, to install the desired couplings between right movers in
wire $b$ to left movers of wire $b+1$, is to adjust the values for
$k_{\zz}$ and $k_{\SO}$ such that all the terms in
$S_{1b}^+S_{1b+1}^-$, except the desired ones, oscillate, \ie
\begin{align}
  \label{eq:4}
  k_{1\R}^{(b)}=-k_{1\L}^{(b+1)}\quad \forall\ n,
\end{align}
while no other values for $k_{1r}$ or $k_{2r}$ match between
neighboring wires. P violation, as defined above, requires the unit
cell to contain more than 2 wires.  A possible choice for $k_{\zz}$
and $k_{\SO}$ within a 3-wire unit cell is given in
Tab.~\ref{tab:microscopic_mom}, and results in the spin flip momenta
$k_{1r}^{(1b)}$ and $k_{2r}^{(1b)}$ of
Tab.~\ref{tab:excitation_mom}. The price we pay for the simplicity of
the construction is that the Zeeman and spin-orbit couplings violate
SU(2) spin symmetry at the single wire level, which is intact in the
KL liquid, but not a required property for the universality class of
CSLs~\cite{yao-07prl247203,gong-15prb075112}.

\begin{table}[]
  \begin{ruledtabular}
    \begin{tabular}{c|c|c} 
      wire number $b$ &$k_{\zz}^{(b)}$ & $k_{\SO}^{(b)}$\\[2pt]\hline 
      $n$ or $n+3$ & $3k_0$ & $k_1$ \\ 
      $n+1$ & $-2k_0$ & $k_0+k_1$ \\ 
      $n+2$ & $-k_0$ & $-2k_0+k_1$
    \end{tabular}
  \end{ruledtabular}
  \caption{Zeeman and spin-orbit momenta in the wires of a unit cell of
    the array.}
  \label{tab:microscopic_mom}
\end{table}

\begin{table}[]
  \begin{ruledtabular}
    \begin{tabular}{c|c|c|c|c} $b$ &$k_{1\R}^{(b)}/2$ &
      $-k_{1\L}^{(b)}/2$&$k_{2\R}^{(b)}/2$ & $-k_{2\L}^{(b)}/2$\\[2pt]\hline 
      $n$   &$3k_0+k_1$  &$-3k_0+k_1$ &$k_\text{F}+k_1$&$-k_\text{F}+k_1$\\ 
      $n+1$ &$-k_0+k_1$  &$3k_0+k_1$  &$k_\text{F}+k_0+k_1$&$-k_\text{F}+k_0+k_1$\\ 
      $n+2$ &$-3k_0+k_1$ &$-k_0+k_1$  &$k_\text{F}-2k_0+k_1$&$-k_\text{F}-2k_0+k_1$
    \end{tabular}
  \end{ruledtabular}
  \caption{Spin-flip excitation momenta associated with the microscopic
    momenta of Tab.~\ref{tab:microscopic_mom} as defined by Eq.~\eqref{eq:k}.}
  \label{tab:excitation_mom}
\end{table}

In terms of bosonic fields, a coupling 
$\frac{1}{2} J (S_{ab}^+S_{a'b+1}^- +\text{H.c.})$
between neighboring wires yields the transverse Hamiltonian
densities
\begin{align} 
  \mathcal{H}_t^{(aa'b)}=2t
  \cos\bigl(\sqrt{2}(\phi_{\s}^{(ab)}-\theta_{\s}^{(ab)}
    +\phi_{\s}^{(a'b+1)}+\theta_{\s}^{(a'b+1)})\bigr)
  \label{eq:chiral_sl_cos}
\end{align}
with 
$2t=J/(2\pi\alpha)^2$.
These terms commute with themselves at different positions $x$ along the
wires (which implies that they can pin the value of the field
combinations forming their argument, and hence open up an energy gap),
and with each other for different values of $b$.
The full Hamiltonian for the wire-coupled Abelian ($k=1$) CSL is hence
given by
\begin{align}
  \label{eq:1}
  H_{k=1}=\sum_{b}\int\! dx
  \left[\mathcal{H}_0^{(1b)}(x) + \mathcal{H}_t^{(11b)}(x)\right].
\end{align}
The state is gapped in the bulk but supports gapless chiral edge modes
in wires $1$ and $N$, which are described by the bosonic fields
$\Phi_1(x)=\bigl(-\phi_{\s}^{(11)}(x)-\theta_{\s}^{(11)}(x)\bigr)/\sqrt{2}$
and $\Phi_N(x)=\bigl(\phi_{\s}^{(1N)}(x)-\theta_{\s}^{(1N)}(x)
\bigr)/\sqrt{2}$.  The corresponding spin flip operators adding spin 1
to the edges can be defined as $S^+_1= \exp(2\i\Phi_{1})$ and $S_N^+=
\exp(2\i\Phi_{N})$. Since the bosonic fields obey the commutation
relations $[\Phi_1(x),\Phi_1(x')]=-(\i\pi/2)\text{sgn}(x-x')$ and
$[\Phi_N(x),\Phi_N(x')]=(\i\pi/2) \text{sgn}(x-x')$, we can identify
the mode $\Phi_1$ ($\Phi_N$) as a left (right) mover with a $K$-matrix
of $K=-2$ ($+2$). This implies half-Fermi (also known as semion)
statistics~\cite{wilczek90,wen04}.

The model further supports gapped bulk excitations described by
$2\pi$-kinks in a sine-Gordon coupling of two neighboring chains
\eqref{eq:chiral_sl_cos}. Since the total spin of the system is given
by
\begin{align}
  S^{\z}_{\text{tot}}
  =&-\frac{1}{\sqrt{2}\pi}\sum_{b}\int\!dx\,
  \partial_x\phi_{\s}^{(1b)}
  =-\frac{1}{2\sqrt{2}\pi}\sum_{b}\nonumber\\
  &\int\!dx\,\partial_x\left(\phi_{\s}^{(1b)}-\theta_{\s}^{(1b)}
    +\phi_{\s}^{(1b+1)}+\theta_{\s}^{(1b+1)}\right)\label{eq:Stot}
\end{align}
modulo edge terms, the spin associated with a kink is $S^\z=1/2$. The
kinks describe spinon excitations, which are fractionalized as the
Hilbert space for a spin 1/2 Mott isolator is spanned by spin flips
operators with \hbox{$S^\z =1$}, which act on a spin polarized vacuum.




\emph{Non-Abelian chiral spin liquids.}---We now consider $k>1$
flavors (or layers) of coupled wires (as illustrated in
Fig.~\ref{fig:array}), and assume $k_{\zz}$ and $k_{\SO}$ as specified
in Tab.~\ref{tab:microscopic_mom} for all flavors.  Spin-spin
couplings between neighboring wires yield three types of cosine terms
of Hamiltonian densities, which do not commute mutually, but preserve
momentum and do commute with themselves at different positions $x$
along the wire.  (In practise, the latter condition implies that we
only need to consider terms which contain two left movers and two
right movers, regardless of whether they stem from creation or
annihilation operators, when we expand four fermion couplings.)
 
The first type is as given in Eq.~\eqref{eq:chiral_sl_cos}, which we
allow for all $a,a'=1,\ldots,k$ with the same coefficient $t$.  (Note
that the commutator between $\mathcal{H}_t^{(aa'b)}(x)$ and
$\mathcal{H}_t^{(cc'b)}(x')$ vanishes only if either $(a,a')=(c,c')$
or $a\ne c \wedge a'\ne c'$.)  The second type is generated by the
coupling $\frac{1}{2} J_{\x\y} (S_{ab}^+S_{a'b}^- + \text{H.c.})$
between different flavors $a$ and $a'$ on the same wire $b$ and takes
the form
\begin{align}
  \mathcal{H}_u^{(aa'b)}
  =2u\cos\bigl(\sqrt{2}(\theta_\s^{(ab)}-\theta_\s^{(a'b)})\bigr),
  \label{u}
\end{align}
with $u=-2J_{\x\y}/(2\pi\alpha)^2$.  (The relative sign between terms
with different Klein factors can be determined by insertion of
$U_{\L\up}U_{\L\dw}U_{\R\up}U_{\R\dw}=1$.)  Finally, the third type is
generated by the coupling $J_{\z} S_{ab}^\z S_{a'b}^\z$ of the
operators
\begin{align}
  S_{ab}^\z
  =&\frac{1}{2}\Bigl( c_{\up}^{(ab)}{}^\dagger c_{\up}^{(ab)}
  - c_{\dw}^{(ab)}{}^\dagger c_{\dw}^{(ab)}\Bigr) \nonumber\\[2pt] 
  =&-\frac{1}{\sqrt{2}\pi}\partial_x\phi_{\s}^{(ab)}\nonumber\\[3pt]
  &+\frac{U_{\L\up}^{(ab)}{}^\dagger U_{\R\up}^{(ab)}}{4\pi\alpha}\,
  e^{-\i2(k_\text{F}+k_{\zz}^{(b)})x}\, 
  e^{\i\sqrt{2}(\phi_\c^{(ab)}+\phi_\s^{(ab)})}+\text{H.c.}\nonumber\\[3pt]
  &-\frac{U_{\L\dw}^{(ab)}{}^\dagger U_{\R\dw}^{(ab)}}{4\pi\alpha}\,
  e^{-\i 2(k_\text{F}-k_{\zz}^{(b)})x}\, 
  e^{\i\sqrt{2}(\phi_\c^{(ab)}-\phi_\s^{(ab)})}+\text{H.c.}
  \label{eq:sz}
\end{align}
between different flavors $a,a'$ on the same wire $b$, and is given
by
\begin{align}
  \mathcal{H}_v^{(aa'b)}
  =2v\cos\bigl(\sqrt{2}(\phi_\s^{(ab)}-\phi_\s^{(a'b)})\bigr),
  \label{v}
\end{align}
with $v=-J_\z/(2\pi\alpha)^2$.  (Note that if the couplings are SU(2)
symmetric, \ie $J_{\x\y}=J_\z$, one finds $u=2v$ in accordance with the energy
density of $S^\x S^\x +S^\y S^\y$ being twice that of $S^\z S^\z$.)

The final Hamiltonian for the non-Abelian CSL at level $k$ is
\begin{align}
  H_{k} =\int\!dx\,\Biggl[
  &\sum_{a,b}\mathcal{H}_0^{(ab)}(x) +
  \sum_{a,a',b}\mathcal{H}_t^{(aa'b)}(x)\, + \nonumber\\
  &\sum_{a<a',b}\Bigl(\mathcal{H}_u^{(aa'b)}(x)+\mathcal{H}_v^{(aa'b)}(x)\Bigr)
  \Biggr]. \label{ham}
\end{align}
A Hamiltonian related to Eq.~\eqref{ham} has been analyzed by Teo and
Kane~\cite{teo-14prb085101} in the context of their CWC of Abelian and
non-Abelian FQH states.  For $k=2$, it yields a Moore-Read
(MR)~\cite{moore-91npb362} phase, and a strong pairing phase of charge
$2e$ bosons at $\nu=1/4$ depending on the parameters $t$, $u$, and
$v$. While the generic problem is intractable due to the
non-commutativity of the different cosine terms, the phase diagram can
still be obtained from the decoupling of individual composite modes in
each wire. (Teo and Kane~\cite{teo-14prb085101} assume the adjustment
of forward scattering terms in their analogous form of Eq.~\eqref{ham}
in such a way that the cosine field arguments decouple at the level of
$\mathcal{H}_0$, and allow for refermionization of their free chiral
spin fields represented by pairs of Majorana fermions. For $k=2$, this
analysis transparently resolves the MR and the strong pairing phase
depending on how the Majorana modes are paired between or within the
wires.) As such, this allows for an effective implementation of the
coset construction in conformal field theory, which then can be used
to yield parafermionic topological phases from the CWC.  In
particular, if we choose the bare coupling parameter $u=v$ (\ie we set
$J_\z=2J_{\x\y}$), the analysis implies that we stabilize a
non-Abelian SU(2) level ${k=2}$ CSL~\cite{greiter-09prl207203}, which
may be viewed as the spin liquid pendant to the MR state. For
arbitrary $k$ and $u = v$ (but regardless of $t$), the same procedure yields a
non-Abelian CSL with level $k$ parafermionic edge modes.  This phase
constitutes the coupled wire pendant of the SU(2)$_k$ non-Abelian
CSL~\cite{greiter-09prl207203}, which, on the level of wave functions,
is obtained from the symmetrization of $k$ Abelian CSLs in the layers.
We should note at this point, however, that an equality of the bare
couplings $u$ and $v$ in Eq.~\eqref{ham} does not guarantee that they
remain equal under the RG flow towards lower energies. Whether the
SU(2)$_k$ parafermionic CSL is the RG fixed point for a given bare
parameter setup depends on the relative coupling strengths of the
cosine terms, and the forward scattering amplitudes.

\emph{Conclusion and Outlook.}---The coupled wire construction of
Abelian and non-Abelian chiral spin liquids offers a deconstructivist
and yet physically motivated, microscopic view on these unconventional
topological quantum states of matter. For the Abelian state, starting
from bosonic spin flip operators which carry spin 1 and couple the
Mott-gaped wires, we have constructed a topological phase with a bulk
gap, fractionalized spin $1/2$ bulk quasiparticles (\ie spinons), and
a single chiral edge mode. We identify this phase with a chiral spin
liquid, and the generalization to multiple layers with SU(2)$_k$
parafermionic chiral spin liquids. The construction outlined above
constitutes the starting point for further study. First, the nature of
the bulk, and in particular the edge, excitations of the SU(2)$_k$
chiral spin liquids require further investigation. Second, a more
rigorous RG treatment of~\eqref{ham} is indispensable in acquiring an
understanding of the phase diagrams of multi-layer Mott-gapped wires
as well as to assess the range of stability for the SU(2)$_k$ chiral
spin liquid states. For $k>2$, this has so far not even been attempted
for the analogous FQHE scenario, and might yield new insights. Third,
from the construction outlined in this Letter, we might also be able
to construct spin liquids without P and T breaking, \ie the spin
liquid pendants~\cite{greiter02jltp1029,scharfenberger-11prb140404} of
a fractional topological insulator~\cite{levin-09prl196803}.

We thank M.~Barkeshli, B.~Bauer, E.~Fradkin, and A.~W.~W. Ludwig for discussions.  This work has been funded by the Helmholtz association through VI-521, the DFG through SFB 1143, and the European Research Council through ERC-StG-TOPOLECTRICS-336012. T.N. acknowledges financial support from DARPA SPAWARSYSCEN Pacific N66001-11-1-4110.

\textit{Note added}: In the final stages of this work, we became aware that a similar idea is being pursued by G.~Gorohovsky, R.~G.~Pereira, and E.~Sela~\cite{goro_note}.

\end{document}